\documentclass[a4paper,fleqn]{cas-dc}

\usepackage[authoryear,longnamesfirst]{natbib}
\usepackage{graphicx}    
\usepackage{amsmath}
\usepackage{siunitx}
\usepackage{circuitikz}
\usepackage{tikz}

\usepackage{soul, xcolor}

\usepackage{algorithm}
\usepackage{algpseudocode}

\def\tsc#1{\csdef{#1}{\textsc{\lowercase{#1}}\xspace}}
\tsc{WGM}
\tsc{QE}

\begin{document}
\let\WriteBookmarks\relax
\def\floatpagepagefraction{1}
\def\textpagefraction{.001}

\shorttitle{Learning Li-ion battery health and
degradation modes from data with
aging-aware circuit models}    

\shortauthors{Z. Zhou, A. Aitio, D. Howey}  

\title [mode = title]{Learning Li-ion battery health and
degradation modes from data with
aging-aware circuit models}  



%

\author[1]{Zihao Zhou}[orcid=0009-0000-1783-8166]



\ead{zihao.zhou@eng.ox.ac.uk}



\affiliation[1]{organization={Department of Engineering Science, University of Oxford},
            city={Oxford},
            postcode={OX1 3PJ}, 
            state={Oxford},
            country={UK}}

\author[1]{Antti Aitio}
\ead{antti.aitio@gmail.com}

\author[1,2]{David Howey}
\cormark[1]

\affiliation[2]{organization={The Faraday Institution},
            city={Didcot},
            postcode={OX11 0RA}, 
            country={UK}}




\cortext[1]{Corresponding author}
\ead{david.howey@eng.ox.ac.uk}


\begin{abstract}
Non-invasive estimation of Li-ion battery state-of-health from operational data is valuable for battery applications, but remains challenging. Pure model-based methods may suffer from inaccuracy and long-term instability of parameter estimates, whereas pure data-driven methods rely heavily on training data quality and quantity, causing lack of generality when extrapolating to unseen cases. We apply an aging-aware equivalent circuit model for health estimation, combining the flexibility of data-driven techniques within a model-based approach. A simplified electrical model with voltage source and resistor incorporates Gaussian process regression to learn capacity fade over time and also the dependence of resistance on operating conditions and time. The approach was validated against two datasets and shown to give accurate performance with less than 1\% relative root mean square error (RMSE) in capacity and less than 2$\%$ mean absolute percentage error (MAPE). Critically, we show that changes from the open circuit voltage versus state-of-charge function will strongly influence the learnt resistance. We use this feature to further estimate \textit{in operando} differential voltage curves from operational data.
\end{abstract}


\begin{highlights}
\item An aging-aware equivalent circuit model is proposed for battery health estimation

\item It estimates capacity and resistance only from operational data without labels

\item Changes of open-circuit voltage are strongly reflected in resistance estimation

\item It may enable an in operando estimation of differential voltage curves

\end{highlights}


\maketitle

\section{Introduction}\label{}
Lithium-ion batteries play an important role in the transition to clean energy. Demand is increasing rapidly for grid storage, electric vehicle and home energy storage applications \citep{tsiropoulos2018li}. One of the most crucial challenges for batteries is estimating their state of health (SOH). Accurate estimation of SOH improves lifetime, system safety, and costs of warranties and maintenance. Many studies \citep{hua2021toward,shahjalal2022review,zhou2020fast} also discuss the importance of  dealing with retired batteries, the number of which will increase drastically in the near future. Estimating the evolution of SOH from historical usage data is crucial for enabling possible `second-life' applications that rely on accurate knowledge of battery residual value.

A commonly used SOH indicator, discharge capacity, normally requires the battery to undergo a low-rate constant-current discharge from 100\% to 0\% state of charge (SOC). This is time consuming, and therefore often not undertaken in real-world applications. Therefore, extensive studies have been done on faster capacity estimation methods \citep{berecibar2016critical,wang2020comprehensive}. These are usually categorised into model-driven and data-driven approaches. Model-driven methods use a specific battery model, such as an electrochemical model or electrical equivalent circuit model (ECM), and update the parameters by repeated fitting as the battery ages  \citep{plett2004extended,gao2021co}.  Electrochemical models usually need half-cell test data for parameterization and tend to suffer from parameter identifiability problems \citep{lu2021implementation, miguel2021review}. Equivalent-circuit models, the focus in this paper, have fewer parameters but are less connected to physical processes. Recursive state estimation techniques, such as extended Kalman filters \citep{wassiliadis2018revisiting} and particle filters \citep{wei2017remaining}, are often adopted for co-estimation of model states and parameters. 

On the other hand, data-driven methods for capacity estimation try to map from carefully selected features to corresponding capacity values \citep{li2019data}. Several features indicative of battery usage and aging may be extracted from cell voltage, current, and temperature measurements \citep{severson2019data, greenbank2021automated, li2024predicting}. Then, machine learning models such as support vector machines \citep{feng2019online}, random forests \citep{li2018random}, neural networks \citep{li2021online}, and Gaussian processes \citep{richardson2018gaussian}, have been used to capture the relationship between these features and capacity. Depending on the choice of inputs, data-driven models can be used either for estimating present SOH, or predicting the future SOH. If the learnt mapping is between input features and future capacity values, these models can be directly used for forecasting. The performance of machine learning models for battery health is largely determined by the breadth, quantity and accuracy of the data used for model training.

Moreover, capacity fade only represents one aspect of battery degradation---increasing impedance (often just represented by resistance) also plays an essential role in battery health \citep{fu2021fast}, leading to power fade, which is often correlated with capacity fade. Impedance values may be easier to estimate from cycling data and field data, where capacity measurements are very challenging \citep{pozzato2023analysis, sulzer2021challenge}. Many SOH estimation works treat impedance as a scalar or rely on a look-up table that needs to be calibrated in standard reference tests \citep{plett2004extended,locorotondo2021development}. However, many experimental works show that impedance does change with operating conditions \citep{waag2013experimental, schonleber2017consistent}. 

In this work, we use an aging-aware electrical equivalent circuit model to estimate both capacity and operando battery impedance (resistance) from cycling data. The model parameters are described as functions of aging time and operating conditions (SOC and applied current in this case) and these functions are learned from data using Gaussian process regression. The model builds on former work \citep{aitio2023learning} but with a simpler ECM emphasizing the importance of the estimated resistance. Results show that the learned resistance function is strongly impacted by the assumptions about the battery open circuit voltage (OCV) versus SOC function. Specifically, aging-induced shape changes in OCV-SOC function are revealed indirectly through the learned resistance function. This leads to a method to estimate differential voltage ($dV/dQ$) curves indirectly from impedance data. Open-source datasets from two batteries with different cycling conditions were used to validate model performance. Our method not only gives accurate performance with less than 1 $\%$ relative RMSE and less than $2\%$ MAPE for capacity estimation, but also enables $dV/dQ$ estimation. Differential voltage analysis enables deeper understanding of battery aging by quantifying metrics such as loss of lithium inventory (LLI) and loss of active material in each electrode  ($\text{LAM}_\text{n}$, $\text{LAM}_\text{p}$) \citep{dubarry2012synthesize, dubarry2022best}. 


\section{Methodology}
\subsection{State space formulation of Gaussian processes}
In this work Gaussian processes (GPs) are used to fit functions to data. Full details of GPs are documented in several books and papers, for example Williams and Rasmussen \cite{williams2006gaussian}---we give only a brief introduction here. A Gaussian process (GP) is defined as a distribution of functions over input $\mathbf{x}$, characterised by a mean and covariance, with the covariance defined by a kernel function $k(\mathbf{x}, \mathbf{x}^{\prime})$,
\begin{equation}
\begin{aligned}
f(\mathbf{x}) & \sim \mathcal{G} \mathcal{P}\left(m(\mathbf{x}), k\left(\mathbf{x}, \mathbf{x}^{\prime}\right)\right). \\
\end{aligned}
\end{equation}

When this input includes both locations in space and time $\mathbf{x} = (s,t)$, it is often called a spatial-temporal Gaussian process \cite{solin2016stochastic},
\begin{equation}
f(s,t) \sim \mathcal{G} \mathcal{P}\left(m(s,t), k\left( (s,t), (s^{\prime}, t^{\prime})\right)\right).
\end{equation}
Without loss of generality, the prior mean function $m(\mathbf{x})$ may be set to zero. By assuming a measurement noise term $\epsilon$, the mapping between $\mathbf{x}$ and $\mathbf{y}$ can be expressed as
\begin{equation}
\mathrm{y}=f(\mathbf{x})+\mathbf{\epsilon}, \mathbf{\epsilon} \sim \mathcal{N}\left(0, \sigma_n^2 \mathbf{I}\right).
\end{equation}
To make predictions, we require the posterior predictive distribution at test samples $\mathbf{X}^{*}$ based on observations from a labeled training set of input-output data $\mathcal{D}=\left\{\left(\mathbf{x}_i, y_i\right)\right\}_{i=1}^{N}$. In this case, the analytical solution is given by \citep{williams2006gaussian},
\begin{equation}
p\left(\mathbf{y}^{*} \mid X^{*}, X, \mathbf{y}\right)=\mathcal{N}\left(\mathbf{y}^{*} \mid \mathbf{m}^{*}, \Sigma^{*}\right), 
\end{equation}
where
\begin{equation}
\begin{aligned}
\label{eq:gp_solution}
\mathbf{m}^* & =\mathbf{K}_{X, *}^T\left[\mathbf{K}_X+\sigma_n^2 \mathbf{I}\right]^{-1} \mathbf{y} \\
\Sigma^* & =\mathbf{K}_{*, *}-\mathbf{K}_{X, *}^T\left[\mathbf{K}_{X}+\sigma_n^2 \mathbf{I}\right]^{-1} \mathbf{K}_{X, *},
\end{aligned}
\end{equation}
and the kernel matrices are simplified as $\mathbf{K}_{X}=k(\mathbf{X}, \mathbf{X})$, $\mathbf{K}_{X,*}=k(\mathbf{X}, \mathbf{X}^{*})$. For a zero-mean GP, the model is defined by its training data and the selected kernel function $k(\mathbf{x}, \mathbf{x}^{\prime})$ including fitted hyperparameters. The negative log maximum likelihood (NLML) estimates of kernel hyperparameters $\mathbf{\theta}$ are given by
\begin{equation}\label{eq:gp_optim}
\begin{aligned}
    &\mathrm{NLML}=-\log p(\mathbf{y} \mid X, \theta)= \\
    & -\frac{1}{2} \mathbf{y}^{\mathrm{T}}\left[\mathbf{K}_{X}+\sigma_n^2 \mathbf{I}\right]^{-1} \mathbf{y}-\frac{1}{2} \log \left|\mathbf{K}_{X} +\sigma_n^2 \right|-\frac{n}{2} \log 2 \pi.
\end{aligned}
\end{equation}

Large datasets pose challenges for GPs due to the requirement for inversion of an 
$N \times N$ matrix (equations \ref{eq:gp_solution} and \ref{eq:gp_optim}) which scales computationally at $\mathcal{O}\left(n^3\right)$. To address this, we adopt a state-space formulation of a GP that instead scales linearly with the `time' dimension, $\mathcal{O}\left(n\right)$.

Proposed by S{\"a}rkk{\"a} et al.\ \citep{sarkka2013spatiotemporal} and Solin \citep{solin2016stochastic}, the state-space formulation interprets a spatial-temporal Gaussian process $f(s,t)$ as the solution to a linear time-invariant stochastic partial differential equation (LTI-SPDE), 
\begin{equation}\label{eq:spde}
\begin{aligned}
\frac{\partial \mathbf{x}(s, t)}{\partial t} & =\mathcal{F} \mathbf{x}(s, t)+\mathbf{L} \omega(s, t) \\
\mathrm{y}_{\mathrm{t}} & =\mathcal{H}_{\mathrm{t}} \mathbf{x}(s, t)+\epsilon_{\mathrm{t}}, \quad \epsilon \sim \mathcal{N}\left(0, \sigma_n^2\right).
\end{aligned}
\end{equation}
We can evaluate this system of equations over a finite collection of spatial points of interest $\{s_i\}_{1}^{n_\text{s}}$. This results in $n_\text{s}$ sets of state vectors, $\left(\mathbf{x}\left(s_1, t\right), \mathbf{x}\left(s_2, t\right), \ldots, \mathbf{x}\left(s_{n_s}, t\right)\right)$, where each is
\begin{equation}
\mathbf{x}\left(s_i, t\right)=\left[f\left(s_i, t\right), \frac{d f\left(s_i, t\right)}{d t}, \ldots, \frac{d f\left(s_i, t\right)^{p-1}}{d t^{p-1}}\right]^{\top}.
\end{equation}
Here, $\mathcal{F}$ and $\mathcal{H}$ are linear operators, $\mathbf{L}$ is a dispersion matrix, and $\omega(\mathrm{x}, t)$ is a spatially resolved white noise process decided by the spatial kernel. The mathematical foundation comes from the Wiener-Khinchin theorem \citep{cohen1998generalization}, which indicates that the GP kernel covariance matrix and the spectral density of a solution process from a LTI-SPDE form a Fourier-transform pair. At a specific spatial point $s_i$, all of the time gradient information up to order $p-1$ is maintained and propagated into future, and this propagation occurs for all $n_\text{s}$ spatial points. The exact value for $p$ is decided by the selected time kernel. Intuitively, one can view this approach as constructing a LTI system which takes a white noise process as input and generates an output that is the desired Gaussian process.

To further simplify the problem, we assume kernel separability,
\begin{equation}
k\left((s, t),\left(s^{\prime}, t^{\prime}\right)\right)=k\left(s, s^{\prime}\right) k\left(t, t^{\prime}\right).
\end{equation}
Under this assumption, $\mathcal{F}$ and $\mathcal{H}$ become constant matrices $\mathbf{F}$ and $\mathbf{H}$. As a result, a GP kernel function may be directly mapped to $\mathbf{F}$, $\mathbf{H}$ and $\omega(s, t)$, defining a corresponding LTI system. The practical implementation of eqn. \ref{eq:spde} requires it to have a discretized form,
\begin{equation}
\begin{aligned}
\mathbf{x}_k & =\mathbf{A}_{k-1} \mathbf{x}_{k-1} +\mathbf{w}_{k-1} \\
\mathrm{y}_k & =\mathbf{H}_k \mathbf{x}_k + v_{k}, \quad v_k \sim \mathcal{N}\left(0, \sigma_n^2\right),
\end{aligned}
\end{equation}
where $\mathbf{x}_k=\left[\mathbf{x}_{s_1,k}, \mathbf{x}_{s_2,k}, \ldots, \mathbf{x}_{s_{n_\text{s}},k}\right]^{\top}$ represents a state vector along all $n_\text{s}$ space locations at time step $t_k$, and $\mathbf{A}_{k}=\exp \left(\Delta t_k \mathbf{I}_{n_\text{s}}\otimes\mathbf{F}\right)$ is a matrix exponential on $\mathbf{F}$ and time interval $\Delta t_k = t_{k+1} - t_k$. 
The Kronecker product here comes from the space discretization. Each space location is propagated in time individually, and correlations in space are captured by the process noise term $\mathbf{w}_{k-1}$ which is a zero-mean multivariate GP over the $n_\text{s}$ space locations with covariance function expressed as,
\begin{equation}
\label{eq:process_noise}
\begin{aligned}
\mathbf{w}_{k-1} &\sim \mathcal{G} \mathcal{P}\left(\mathbf{0}, \mathbf{K}_\text{s} \otimes w_{k-1}\right)\\
w_{k-1}&=\int_0^{\Delta t_k} \mathbf{A}_{k-1} \mathbf{L} \mathrm{Q}_{\mathrm{c}} \mathbf{L}^{\top} \mathbf{A}_{k-1}^{\top} \mathrm{d} \tau.
\end{aligned}
\end{equation}
Here $\mathbf{K}_s$ is the $n_\text{s} \times n_\text{s}$ covariance matrix of the spatial kernel $k(s,s^{\prime})$ at $n_\text{s}$ locations, and spectral density matrix $\mathrm{Q}_\text{c}$ is decided by the temporal kernel $k(t,t^{\prime})$ as a mathematical combination of kernel hyperparameters. 

Once the GP is formulated as a discrete-time LTI system, both the estimation of posterior predictive distribution in eqn.\ \ref{eq:gp_solution} and the hyperparameter NLML in eqn.\ \ref{eq:gp_optim} can be achieved recursively using a Kalman filter and Rauch-Tung-Striebel (RTS) smoother \citep{sarkka2023bayesian}, enabling linear time complexity in forward model runs \citep{aitio2023learning}. 

\subsection{Gaussian processes for equivalent circuit models}

\begin{figure}
\centering
\begin{circuitikz}[european]
    \draw(0,0) to[vsource, l=$V_0(z)$] (0,2) 
    to[R,l=$R_0(z{,}I{,}\zeta)$] (4,2)  
    to[short, -, i=$I$] (3,2)
    (4,2) to[short, o-] (4,2)
    (4,0) to[short, o-] (0,0)
    (4,0) to[open, v=$V_\text{term}$, o-o] (4,2);
\end{circuitikz}
\caption{Equivalent circuit model for Li-ion battery}
\label{fig:ecm}
\end{figure}
As shown in Fig.\ \ref{fig:ecm}, a simple ECM is used for modelling the battery dynamics in this work. This approach is similar to the method we used in Aitio et al.\ \cite{aitio2023learning} but simplified. The corresponding mathematical formulation is given as
\begin{equation}\label{eq:ecm}
\begin{aligned}
V_\text{term}&=V_0\left(z\right)+R_0\left(z, I,\zeta\right) I\\
\frac{\mathrm{d} z}{\mathrm{d} t}&=I Q^{-1}\left(\zeta\right)
\end{aligned}
\end{equation}
where $z$ is the SOC, $I$ is the applied current, and $V_0(z)$ represents the OCV-SOC curve, which is often measured at beginning of life and assumed to be fixed over life. Two timescales are included in this ECM, one is the operating timescale ($t$), which is often at the seconds level. Another is the life-long aging timescale ($\zeta$), which can vary from days to months depending on usage. Although this circuit model is simple, for a high quality commercial cell operated at low rates, the internal overpotentials should be small, and therefore this model produces reasonable results. Even in highly dynamic applications, there are usually still some periods of low-rate operation that could be used for health estimation.

The variables $I$, $z$, $V_\text{term}$ are all functions of $t$, unless written in their discrete-time versions that are demarcated with subscripts (e.g.\ $z_i$). For clarity the time-dependence of the continuous-time variables is assumed, rather than written out in every equation. For simplicity, we use a vector to denote operating conditions,
\begin{equation}
    \mathbf{s} = [z, I]^{\top}.
\end{equation}
The inverse capacity term $Q^{-1}$ is modeled as a GP over $\zeta$, while the resistance term $R_0$ is modeled as a GP over $\mathbf{s}$ and $\zeta$,
\begin{equation}
\label{eq:QR_GP}
\begin{aligned}
Q^{-1}(\zeta) & \sim \mathcal{G} \mathcal{P}\left(m(\zeta), k\left(\zeta, \zeta^{\prime}\right)\right) \\
R_0\left(\mathbf{s},\zeta\right) & \sim \mathcal{G} \mathcal{P}\left(m\left(\mathbf{s},\zeta\right), k\left(\left(\mathbf{s},\zeta\right),\left(\mathbf{s}^{\prime},\zeta^{\prime}\right)\right)\right).
\end{aligned}
\end{equation}
Temperature was tightly controlled and consistent for battery samples used in our work, so had little impact on the results. Temperature effects can however be included as another dimension of the operating-condition dependency of $R_0(s,\zeta)$, where $s=[z,I,T]$ \cite{aitio2021predicting}. Given the nonzero nature of impedance and capacity, instead of directly modelling $R_0$ and $Q$, we chose to model them each as affine transformations of a zero-mean GP, which is equivalent to setting a nonzero prior mean for impedance and capacity \cite{aitio2023learning}. Specifically, Eqns.\ \ref{eq:QR_GP} are written as, 
\begin{equation}
\begin{aligned}
Q^{-1}(\zeta) & =q_0(1+q(\zeta)) \\
R_0\left(\mathbf{s}, \zeta\right) & =r_0\left(1+r\left(\mathbf{s}, \zeta\right)\right) \\
q(\zeta) & \sim \mathcal{G} \mathcal{P}\left(0, k\left(\zeta, \zeta^{\prime}\right)\right) \\
r\left(\mathbf{s}, \zeta\right) & \sim \mathcal{G} \mathcal{P}\left(0, k\left(\left(\mathbf{s}, \zeta\right),\left(\mathbf{s}^{\prime}, \zeta^{\prime}\right)\right)\right)
\end{aligned}
\end{equation}
where constants $r_0,q_0$ reflect prior assumptions for the values of $R_0, Q^{-1}$, respectively. In practice, these can be set to the corresponding beginning-of-life (BoL) values. As a Gaussian distribution remains Gaussian under arbitrary affine transformation, $R_0$ and $Q^{-1}$ are still GPs.

The operating-condition dependency on $\mathbf{s}$ and the life-long aging dependency on $\zeta$ are modelled with different kernels, since we expect that degradation is irreversible, whereas a more general smooth function describes operating-condition dependency,
\begin{equation}
k\left(\left(\mathbf{s}, \zeta\right),\left(\mathbf{s}^{\prime}, \zeta^{\prime}\right)\right)=k_{\text {Matern32 }}\left(\mathbf{s}, \mathbf{s}^{\prime}\right) k_{\mathrm{WV}}\left(\zeta, \zeta^{\prime}\right)
\end{equation}
where $k_{\mathrm{WV}}$ and $k_{\mathrm{Matern32}}$ represent a Wiener velocity (WV) kernel and a Matern ($\nu=3/2$) kernel, respectively, so
\begin{equation}
\begin{aligned}
& k_{\mathrm{WV}}\left(\zeta, \zeta^{\prime}\right)=\sigma_\zeta^2\left(\frac{\min ^3\left(\zeta, \zeta^{\prime}\right)}{3}+\left|\zeta-\zeta^{\prime}\right| \frac{\min ^2\left(\zeta, \zeta^{\prime}\right)}{2}\right) \\
& k_{\text {Matern32 }}\left(\mathbf{s}, \mathbf{s}^{\prime}\right)=\sigma_s^2(1+\sqrt{3} d) \exp (-\sqrt{3} d) \\
& d^2=\left|\mathbf{s}-\mathbf{s}^\prime\right|^{\top}\left[\begin{array}{cc}
l_z^{-2} & 0 \\
0 & l_I^{-2}
\end{array}\right]\left|\mathbf{s} - \mathbf{s}^\prime\right|.
\end{aligned}
\end{equation}
While stationary kernels eventually converge back to the mean when extrapolated, the WV kernel is a non-stationary kernel that is more suitable for extrapolating aging trends into the future.

\begin{figure*}
\begin{center}
\includegraphics[width=16cm]{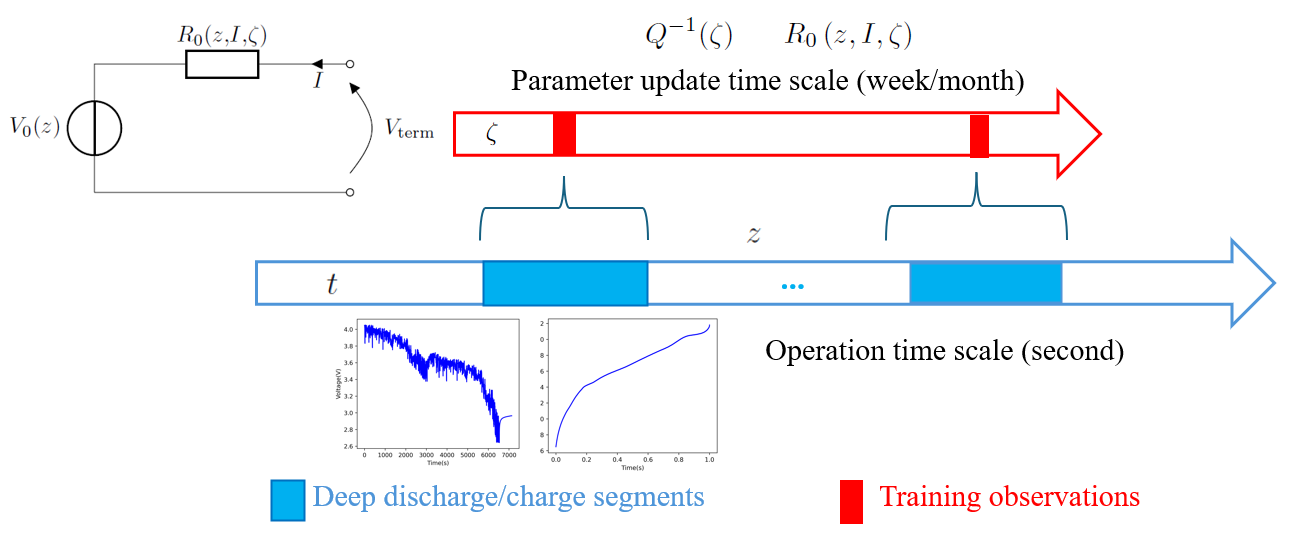}    
\caption{Modelling the separation of timescales with battery operating dynamics that change quickly (as a function of $t$) vs.\ aging dynamics that change slowly (as a function of $\zeta$)} 
\label{fig:twoscale_logic}
\end{center}
\end{figure*}

As shown in Fig.\ \ref{fig:twoscale_logic}, the battery operating dynamics, described by the ECM, and aging dynamics, described by the GPs, are evaluated at two different timescales. While the ECM state ($z$) changes over seconds during cycling ($t$), the GPs representing $Q^{-1}$ and $R_0$ may change more slowly, over weeks or months, as a function of cell-aging timescale $\zeta$. In subsequent sections we will use discrete-time rather than continuous-time dynamics, with the indexes $i$ and $j$ denoting the discrete-time variables associated with $t$ and $\zeta$, respectively, with the faster discrete time step size defined as $\Delta t_i$ and the slower step size as $\Delta \zeta_j$.

The discrete-time state propagation for the ECM is a simple Coulomb counter that can be expressed as
\begin{equation}
z_i=z_{i-1} + Q^{-1} I_i \Delta t_i .
\end{equation}
The discrete-time propagation for the inverse capacity and resistance parameters, which are GPs, can be written as
\begin{equation}
\label{eq:gp_state_propagate}
\begin{aligned}
\mathbf{x}_{Q^{-1}, j}&=\exp \left(\Delta \zeta_j \mathbf{F}_{\mathrm{WV}}\right) \mathbf{x}_{Q^{-1}, j-1} +w_{j-1}\\
\mathbf{x}_{R_0, j}&=\exp \left(\Delta \zeta_j \mathbf{I}_{n_\text{s}} \otimes \mathbf{F}_{\mathrm{WV}}\right) \mathbf{x}_{R_0, j-1}+\mathbf{w}_{j-1},
\end{aligned}
\end{equation}
where, for the WV kernel, $\mathbf{F}_{\mathrm{WV}} = \left[\begin{array}{ll}
0 & 1 \\
0 & 0
\end{array}\right]$ represents the discrete-time interval on the lifelong aging scale. Because of the operating condition dependency of $R_0$, Kronecker products are used to describe the corresponding discretisation in $\mathbf{I}_{n_\text{s}} \otimes \mathbf{F}_{\mathrm{WV}}$ and $\mathbf{w}_{j-1}$. In this work, $n_\text{s} = n_z n_I$. Specifically, the 0 to 1 SOC range is evenly divided into $n_z$ discrete SOC levels, while the applied current range (in absolute scale) is evenly divided into $n_I$ discrete current levels. The process noise terms $w_{j-1}$ and $\mathbf{w}_{j-1}$ can be calculated following Eqn.\ \ref{eq:process_noise}, where $\mathrm{Q}_c$ for the WV kernel is $\sigma_{\zeta}^{2}$.

\subsection{Joint estimation of battery states and GPs}
The state covariance initialisation and process noise matrix of the joint system and GP subsystem are summarized as follows:
\begin{equation}
\begin{aligned}
& \mathbf{P}_{\text{joint},0}=\left[\begin{array}{cc}
\mathrm{P}_{z, 0} & 0  \\
0 & \mathbf{P}_{\text{GP}, 0}
\end{array}\right], \mathbf{P}_{\mathrm{G P}, 0}=\left[\begin{array}{cc}
\mathbf{P}_{Q^{-1}, 0} & 0 \\
0 & \mathbf{P}_{R_0, 0}
\end{array}\right] \\
& \mathbf{W}_{\text{joint}}=\left[\begin{array}{cc}
\mathrm{W}_{z} & 0  \\
0 & \mathbf{W}_{\text{GP}}
\end{array}\right], \mathbf{W}_{\mathrm{G P}}=\left[\begin{array}{cc}
\mathbf{W}_{Q^{-1}} & 0 \\
0 & \mathbf{W}_{R_0},
\end{array}\right] 
\end{aligned}
\end{equation}
where
\begin{equation}
\begin{aligned}
&\mathbf{P}_{R_0, 0}=\mathbf{K}_{\mathrm{Mat}} \otimes \mathbf{P}_{\zeta_0, \mathrm{WV}}, \quad\mathbf{P}_{Q^{-1}, 0}=\mathbf{P}_{\zeta_0, \mathrm{WV}}\\
&\mathbf{W}_{R_0}=\mathbf{K}_{\mathrm{Mat}} \otimes \mathbf{W}_{\mathrm{WV}}, \quad\mathbf{W}_{Q^{-1}}=\mathbf{W}_{\mathrm{WV}}.
\end{aligned}
\end{equation}
In these equations, $\zeta_0$ represents the initial aging time step and  $\mathbf{K}_{\mathrm{Mat}}$ is the covariance matrix of the Matern kernel at $n_s$ discrete operating conditions. For the selected WV kernel, the initial covariance and noise matrix are   
\begin{equation}
\mathbf{P}_{\zeta_0, \mathrm{WV}}=\sigma_{\zeta}^2\left[\begin{array}{cc}
\frac{1}{3} \zeta_0^3 & \frac{1}{2} \zeta_0^2 \\
\frac{1}{2} \zeta_0^2 & \zeta_0
\end{array}\right], \mathbf{W}_{ \mathrm{WV}}=\sigma_{\zeta}^2\left[\begin{array}{cc}
\frac{1}{3} \Delta \zeta^3 & \frac{1}{2} \Delta \zeta^2 \\
\frac{1}{2} \Delta \zeta^2 & \Delta \zeta
\end{array}\right].
\end{equation}

In literature, several works \citep{hu2012multiscale,wassiliadis2018revisiting} use a dual estimation framework to propagate these two subsystems separately, with the estimated SOC ($z$) as the `observation' for parameter updates. By decoupling states from parameters, however, any cross-correlations between the two are lost and this may lead to poor estimation accuracy \citep{plett2015battery}. For example, the uncertainty information from the SOC estimation, which is important for parameter updates, is discarded. As an alternative, other works \citep{liu2020joint,yu2017lithium} adopt a joint estimation framework with a single timescale to update both states and parameters. This may suffer from poor numerical conditioning, though, due to the vastly different timescales of the state and parameter dynamics (seconds versus months).

To address these issues, we adopted our previously proposed co-estimation framework  \citep{aitio2021predicting}. Within charge or discharge segments, the GP states (i.e.\ circuit model parameters) and the ECM state are propagated jointly. This means the GP dynamics described in eqn.\ \ref{eq:gp_state_propagate} also evolve within each segment over time steps of $\Delta t$. However, since $\Delta t \ll \Delta \zeta$, the changes in $Q$ and $R_0$ over small timesteps will be small since the GP is relatively smooth on the timescale of $\Delta t$. Between segments, only the GP states are preserved; the ECM state is reinitialized at the beginning of every new discharge segment based on the measured initial SOC conditions. In this way, the GPs describing capacity and resistance and the ECM describing state of charge dynamics each share the same observations (eqn.\ \ref{eq:ecm}), but their propagation is at different timescales.

Within a segment, the overall joint state vector for both ECM and GPs is given by
\begin{equation}
\mathbf{x}_{\text {joint}, i}=\left[\begin{array}{c}
z_i \\
\mathbf{x}_{Q^{-1}, i} \\
\mathbf{x}_{R_0, i}
\end{array}\right].
\end{equation}
Between segments, the GP subsystem is linear and the GP state vector is given by
\begin{equation}
\mathbf{x}_{\text{GP}, j}=\left[\begin{array}{l}
\mathbf{x}_{Q^{-1}, j} \\
\mathbf{x}_{R_0, j}
\end{array}\right].
\end{equation}
Note that within a segment the joint system is nonlinear because Eqn.\ \ref{eq:ecm} is nonlinear in SOC ($z$) due to the $V_0(z)$ and $R_0(z)$ functions. For simplicity, we write the nonlinear joint system dynamics as
\begin{equation}
\label{eq:joint_sys}
\begin{aligned}
\mathbf{x}_{\text {joint}, i} & =g\left(\mathbf{x}_{\text {joint},i-1}, I_{i-1}, \mathbf{W}_{\text {joint}, i-1}\right) \\
V_i & =h\left(\mathbf{x}_{\text {joint}, i}, I_{i-1}, v_i\right).
\end{aligned}
\end{equation}
where $v_i$ represents the measurement noise with variance $\sigma^{2}_v$. An extended Kalman filter is used to solve the system dynamics with local linearisation via Jacobian matrices
\begin{equation}
\mathbf{G}_i=\frac{\mathrm{d} g}{\mathrm{d} \mathbf{x}_{\text {joint},i}}, \quad \mathbf{H}_i=\frac{\mathrm{d} h}{\mathrm{d} \mathbf{x}_{\text{joint},i}}.
\end{equation}

The full co-estimation framework is given in Algorithm 1. Because $R_0$ is modelled with a GP, its predictive variance needs to be included when calculating the noise covariance of output (the $I_i^{2}\mathbf{\Sigma}_{R_0,i}$ term in $\mathbf{S}_i$). Detailed derivations can be found in Appendix \ref{appenA} and \ref{appenB}. To provide a baseline comparison against our method, we also implemented a random-walk dual-estimation approach for parameter updates, see Appendix \ref{appenD}.

The co-estimation in Algorithm 1 can be viewed as a single forward run of the proposed model under a specific GP hyperparameter set and measurement noise variance (i.e., $\sigma_\zeta^2, \sigma_s^2, l_z^{-1}, l_I^{-1}, \sigma_v^2$). To find the optimal values of these, we use the accumulated NLML ($\Phi$) as our cost function. Because of the additive nature of the NLML, $\Phi$ can be updated during the state and parameter co-estimation process. An outer optimisation loop using a gradient-based L-BFGS-B algorithm is then used to obtain \textit{maximum a posteriori} (MAP) estimates of the hyperparameters for GPs, running the forward model many times, with each forward run giving one evaluation of $\Phi$. The hyperparameter optimisations were solved on a virtualized Linux platform with two physical Intel Xeon Silver 5216 CPUs at 2.1 GHz. Finding the optimal hyperparameters took approximately 60 and 90 minutes for cells A and B, respectively.

\begin{algorithm}
   \caption{Co-estimation of battery state and GPs using extended Kalman filter. The NLML ($\Phi$) is calculated recursively on every terminal voltage measurement.}
    \begin{algorithmic}[1]
      \State \textbf{Initialisation of co-estimation at} $\zeta =\zeta_0$

        \State $\mathbf{x}_{\mathrm{GP}}^{+}=\mathbf{x}_{\mathrm{GP}, 0}, \mathbf{P}_{\mathrm{GP}}^{+}=\mathbf{P}_{\mathrm{GP}, 0}$
        \State $\Phi=0$
        \For{$\text{seg}_{j} \in \text{segments}$}
            \State \textbf{GP propagation}
            \State $\mathbf{x}_{\mathrm{GP},j}^{-}=\exp (\mathbf{F} \Delta \zeta_j) \mathbf{x}_{\mathrm{GP},j-1}^{+}$
            \State $\mathbf{P}_{\mathrm{GP},j}^{-}=\exp (\mathbf{F} \Delta \zeta_j) \mathbf{P}_{\mathrm{GP}}^{+} \exp (\mathbf{F} \Delta \zeta_j)^{\mathrm{T}}+\mathbf{W}_{\mathrm{GP}}(\Delta \zeta_j)$
            \State \textbf{Initialisation at start of $j$th segment}
            \State $\mathbf{x}_{\text{joint}}^{+}=\mathbf{x}_{\text{joint},0} = [z_0, \mathbf{x}_{\text{GP},j}^{-}]^{\top}$
            \State $\mathbf{P}_{\text{joint}}^{+}=\mathbf{P}_{\text {joint}, 0}=\left[\begin{array}{cc}
            \mathbf{P}_{z, 0} & 0 \\
            0 & \mathbf{P}_{\mathrm{GP},j}^{-}
            \end{array}\right]$
            \For{$V_i \in \text{seg}_j$}
                \State \textbf{Joint states propagation}
                \State $\mathbf{x}_{\text {joint}, i}^{-}=g\left(\mathbf{x}_{\text {joint},i-1}^{+}, I_{i-1}, \mathbf{W}_{\text {joint}, i-1}\right)$
                \State $\mathbf{P}_{\text{joint},i}^{-}=\mathbf{G}_{i-1} \mathbf{P}_{\text{joint},i-1}^{+} \mathbf{G}_{i-1}^{\mathrm{T}}+\mathbf{W}_{\text{joint},i-1}$
                \State \textbf{Joint states update}
                \State $\mathbf{e}_{i}=V_i-h\left(\mathbf{x}_{\text {joint}, i}^{-}, I_{i-1}, v_i\right)$
                \State $\mathbf{S}_{i}=\mathbf{H}_{i} \mathbf{P}_{\text{joint},i}^{-} \mathbf{H}_{i}^{\top}+ I_{i}^{2}\mathbf{\Sigma}_{R_0, i} + \sigma_{v}^2$
                \State $\mathbf{L}_{i}=\mathbf{P}_{\text{joint},i}^{-} \mathbf{H}_{i}^{\top} \mathbf{S}_{i}^{-1}$
                \State $\mathbf{x}_{\text{joint},i}^{+}=\mathbf{x}_{\text{joint},i}^{-}+\mathbf{L}_{i} \mathbf{e}_{i}$
                \State $\mathbf{P}_{\text{joint},i}^{+} = (\mathbf{I}-\mathbf{L}_i \mathbf{H}_{i})^{\top}\mathbf{P}_{\text{joint},i}^{-}(\mathbf{I}-\mathbf{L}_i \mathbf{H}_i)+\mathbf{L}_{i}\mathbf{e}_i\mathbf{L}_{i}^{\top}$
                \State $\Phi=\Phi + \frac{1}{2} \mathbf{e}_{i}^{\top} \mathbf{S}_{i}^{-1} \mathbf{e}_{i}+\frac{1}{2} \log \left|2 \pi \mathbf{S}_{i}\right|$
        \EndFor
        \EndFor
\end{algorithmic}
\end{algorithm}

\section{Data preprocessing and model setup}
\subsection{Data description}
\subsubsection{Cell A}
Cell A is a nickel-manganese-cobalt vs.\ graphite Li-ion battery from an open test dataset \citep{opendataset}. It is a commercial 502030 cell with nominal capacity \SI{250}{mAh} and operating voltage from 3.0 to \SI{4.2}{V}. The true initial capacity was measured as around \SI{280}{mAh} and the beginning-of-life DC resistance around \SI{130}{m$\Omega$}. This battery was aged with repeated charge/discharge cycles and checked through a weekly reference performance test (RPT). Cycling aging was conducted with a constant \SI{1.825}{C} charging C-rate and a constant \SI{0.5}{C} discharging C-rate. The depth of discharge (DoD) was $100\%$ and the environmental temperature was kept at \SI{30}{\celsius}. The RPT consisted of low rate (\SI{0.2}{C}) constant-current constant-voltage (CC-CV) charging and constant-current discharging, yielding ground truth discharge capacity measurements and an up-to-date pseudo OCV-SOC curve. The cell number is G37C2 in the original dataset.

\subsubsection{Cell B}
Cell B is a nickel-cobalt-aluminium oxide vs.\ graphite-silicon battery from another open dataset \citep{jost2021timeseries}, with cell ID 015. It is a commercial 18650 cell with nominal capacity \SI{3400}{mAh} and BoL DC resistance around \SI{25}{m$\Omega$}. This battery was charged with a CC-CV regime and discharged using a scaled drive cycle profile from 0--\SI{9.7}{A} (absolute value). Cycle aging was conducted at \SI{25}{\celsius} between $10\%$ and $90\%$ SOC. Regular RPTs were taken every 30 cycles, at \SI{0.1}{C}. 

\subsection{Extracting discharge segments from cycling data}

The sampling frequency for cycling was \SI{0.2}{Hz} and \SI{1}{Hz} for cells A and B, respectively. The corresponding full cycling datasets (over all of life) are 1,840,632 and 41,346,756 rows of current and voltage measurements, respectively. Using the entire cycle life data to optimize GP hyperparameters requires substantial computational effort. However, it is reasonable to assume that the aging between two adjacent cycles is small, and therefore we only selected a small number of cycling segments from which to train the model, with relatively large between-segment distances $\Delta \zeta_j$. 

In this work, the process of down-selecting discharge segments was as follows. Discharge segments were selected evenly over the life of each battery. In the case of cell A, 15 discharge segments in total were chosen, and in the case of cell B, 18 segments were chosen. This reduced the data volume to 21,881 and 120,801 rows, respectively. In practice, the time interval $\Delta \zeta_j$ between segments can be in the unit of days, weeks or months depending on usage and test conditions. The method would also work using units such as equivalent full cycles or cumulative charge throughput. One reason for using days is that segment extraction can be done in a convenient manner. The first 10 segments were used to train the model (i.e.\ to optimize the kernel hyperparameters for the GPs). We chose 10 segments for model training since it was computationally efficient but sufficient to demonstrate the efficacy of our method. 

\begin{figure*}
\centering
\includegraphics[width=12cm]{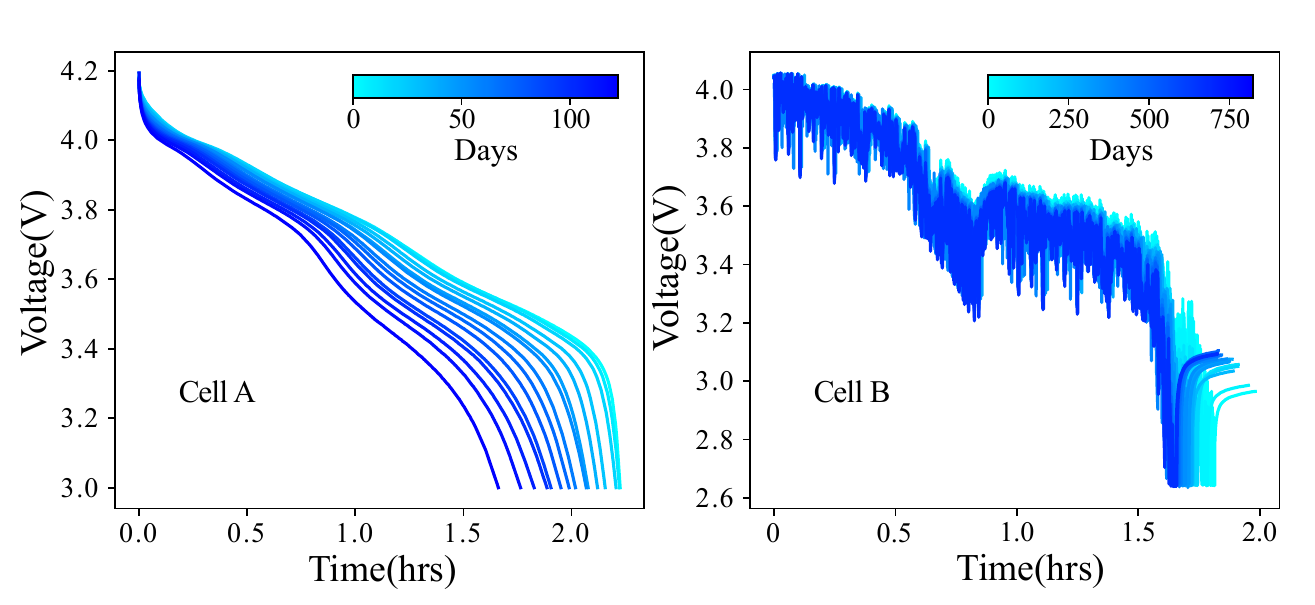}  
\caption{All extracted discharge segments, cells A and B} 
\label{fig:dis_segments}
\end{figure*}

The RPT data at BoL was used to calibrate the OCV-SOC relationship, while other RPT data were not used in the proposed model. The extracted discharge segments for cells A and B are shown in Fig.\ \ref{fig:dis_segments}. The aging effect can be clearly seen from the shrinkage of discharge curves on the time axis.

Considering the computational cost and data range, we set the operating condition discretization for cell A to $n_z=25$ and $n_I=1$, and for cell B to $n_z=25$ and $n_I=5$.

\section{Results and Discussion}

\subsection{Health estimation and prediction}
\begin{figure}
\begin{center}
\includegraphics[width=8cm]{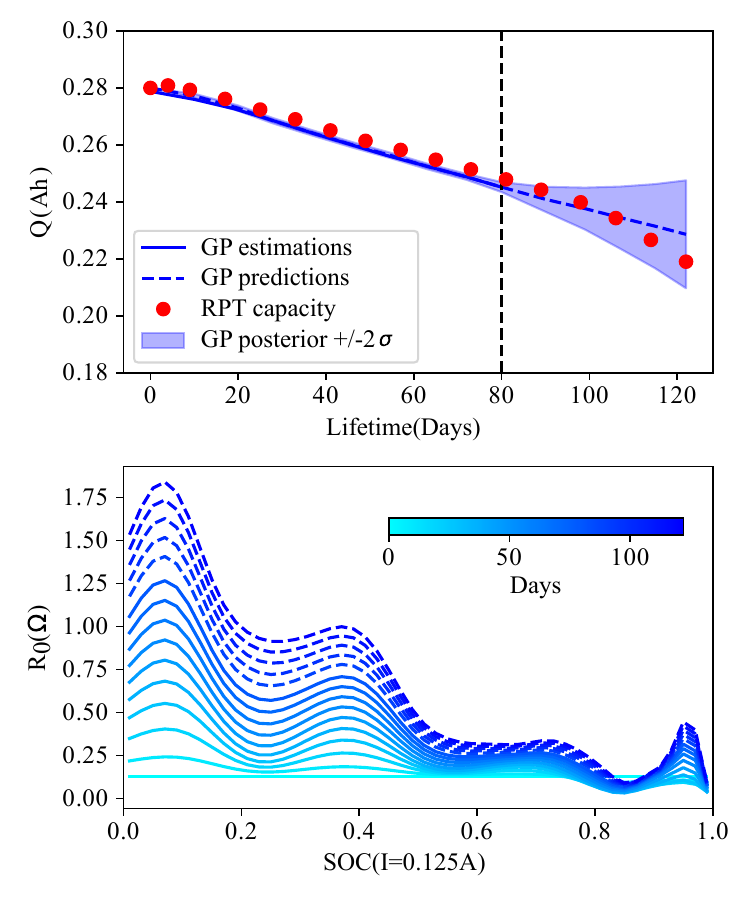} 
\caption{Cell A projections of GP posteriors for $Q$ and $R_0$. Ground truth RPTs are only used for capacity validation (red dots). RMSE and MAPE for capacity estimates are \SI{0.0026}{Ah} and 0.86$\%$, and for predictions are \SI{0.0038}{Ah} and 1.53$\%$} 
\label{fig:CellA_result}
\end{center}
\end{figure}

\begin{figure}
\begin{center}
\includegraphics[width=8cm]{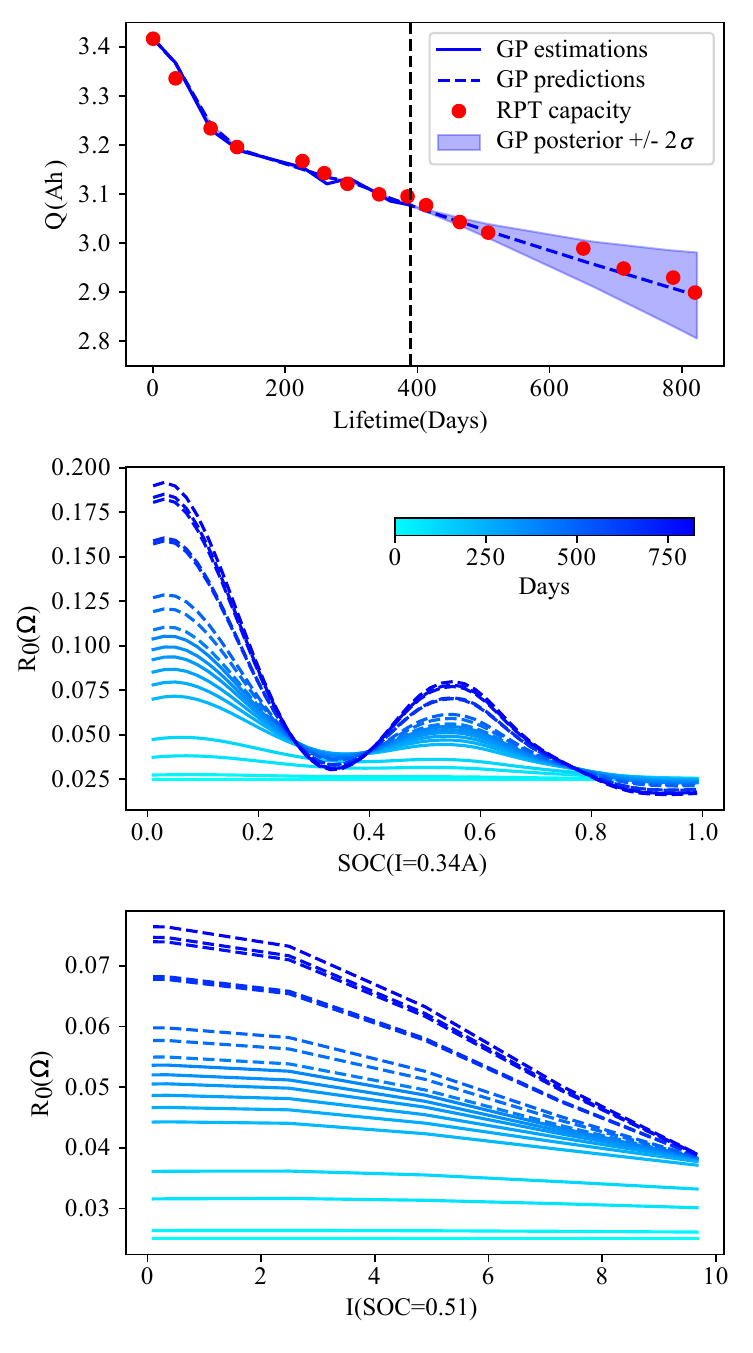} 
\caption{Cell B projections of GP posteriors for $Q$ and $R_0$. Ground truth RPTs  are only used as capacity validations (red dots). RMSE and MAPE for capacity estimates are \SI{0.0122}{Ah} and 0.31$\%$, and for predictions are \SI{0.009}{Ah} and 0.27$\%$}
\label{fig:CellB_result}
\end{center}
\end{figure}

Fig.\ \ref{fig:CellA_result} and Fig.\ \ref{fig:CellB_result} show projections of the GP estimates of discharge capacity $Q$ and impedance $R_0$ over battery lifetime for cells A and B, respectively. Comparisons against the baseline random-walk method (Appendix \ref{appenD}) show that GP estimates are much more accurate and smoother than the random-walk approach. The train-test split is shown by the vertical black line. Here, `training' refers to the data used to fit GP hyper-parameters, whereas `testing' refers to short-term future predictions that are obtained by running the parameter dynamics forward in time without seeing new data. Such short-term future SOH predictions are useful for many applications, e.g., battery early maintenance. For cell A, the discharge current is constant, thus $R_0$ is only plotted against SOC. For cell B, with dynamic discharging profile, $R_0$ is plotted against both SOC and applied current.

It is clear that the GP captures the aging trend for capacity fade well based only on cycling data without RPTs. As the non-stationary WV kernel is used for modelling the time dependency of $Q^{-1}(\zeta)$, the GP extrapolation maintains the last known aging trend from the training stage. This linear extrapolation fits well for cell B, while it appears slightly worse for cell A where the ground truth aging speed accelerates in later life. The uncertainty for estimates is shown by the shaded area that grows larger when extrapolating further into the future.

For the estimation results of $R_0$, a clear upwards shift can be found for cell A during aging, indicating a systematic increase of battery resistance over all SOC levels. For cell B, $R_0$ changes in a more complex way, with both increasing and decreasing trends observed against different SOC levels. The scale of $R_0$ dependency on applied current is much smaller than the SOC dependency. Thus, we mainly discuss the SOC dependency in this work, and more details about the applied current dependency can be found in Appendix \ref{appenC}. 

All estimations are shown in solid lines, extrapolations are shown as dotted lines. The blue gradient colors denote the corresponding aging times. The extrapolation for $R_0$ also maintains the increasing trend of training stage. To keep the figure simple, the uncertainties for $R_0$ estimations are included in Fig.\ \ref{fig:R0_increase}.

In order to ensure smoothness in the estimated $R_0$ functions that are displayed, the discretization of SOC was further increased to 100 evenly spaced locations. The trained model was then rerun in forward mode, with the previously learnt GP hyperparameters, but using the extended SOC discretization.

\subsection{Dependence of $R_0$ on SOC}

\begin{figure*}
\begin{center}
\includegraphics[width=12cm]{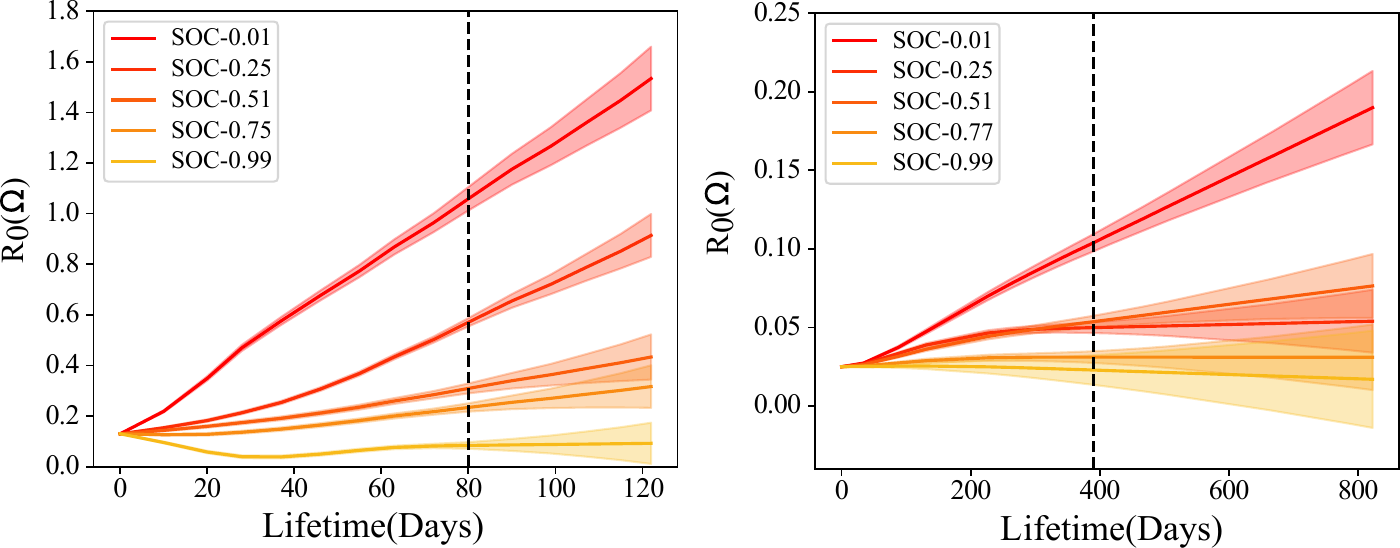}  
\caption{GP estimation of resistance increase at different SOC levels} 
\label{fig:R0_increase}
\end{center}
\end{figure*}

Considering the results of the previous section, there are two abnormal features to note in the evolution of $R_0(z, \zeta)$. First, the increasing scale of $R_0$ is much larger than expected. Given that the BoL DC resistance calculated from a pulse test is around \SI{0.13}{$\Omega$} and \SI{0.025}{$\Omega$} for cells A and B respectively, the increase in $R_0$ by nearly five times at low SOC level as the cell ages seems unrealistic. Second, there is a wave-like curve shape and a  corresponding uneven increase of resistance against SOC. This SOC dependency is also observed by others \citep{waag2013experimental, schonleber2017consistent}. Fig.\ \ref{fig:R0_increase} shows impedance increases at different SOC levels. In the training set, the increase at high SOC ($>0.9$) is almost negligible for both cells. However, at low SOC ($<0.1$), the increase of $R_0$ is nearly \SI{0.88}{$\Omega$} and \SI{0.07}{$\Omega$} for cell A and B respectively. 

We propose that these unexpected behaviours are not caused by `true' changes in the underlying resistance, but instead hypothesize that a significant reason for increases in the estimated $R_0(z,\zeta)$ is shape changes in the battery OCV-SOC curve as it ages. 
Specifically, the battery dynamics in Eqn.\ \ref{eq:ecm} assume that the OCV-SOC functional relationship is kept the same across the entire  life, and only scaled by $Q$. A more realistic formulation could be expressed as
\begin{equation}
V_\text{term}=V_0(z, \zeta)+\hat{R}_0(z, \zeta) I,
\end{equation}
where the OCV-SOC relationship $V_0(z, \zeta)$ is itself a function of both SOC ($z$) and aging lifetime ($\zeta$). In this case, $\hat{R}_0$ is the `pure' impedance term (uneffected by OCV). By further separating the OCV term into a fixed part $\bar{V}_0(z, \zeta_0)$ (representing BoL OCV-SOC), and an age-sensitive variable part $\Delta V_0(z, \zeta)$, one can rewrite this equation as
\begin{equation}
\begin{aligned}
& V_\text{term}=\bar{V}_0\left(z, \zeta_0\right)+\Delta V_0(z, \zeta)+\hat{R}_0(z, \zeta) I \\
& V_\text{term}=\bar{V}_0(z)+I \underbrace{[\left. \frac{\Delta V_0(z, \zeta)}{I} +\hat{R}_0(z, \zeta)\right]}_{R_0(z, \zeta)}.
\end{aligned}
\end{equation}
It is now clear that, rather than obtaining the pure resistance $\hat{R}_0$, the learnt  function $R_0(z, \zeta)$ is influenced both by changes in the resistance and changes in the OCV-SOC functional form as the battery ages. 

\begin{figure*}
\begin{center}
\includegraphics[width=16cm]{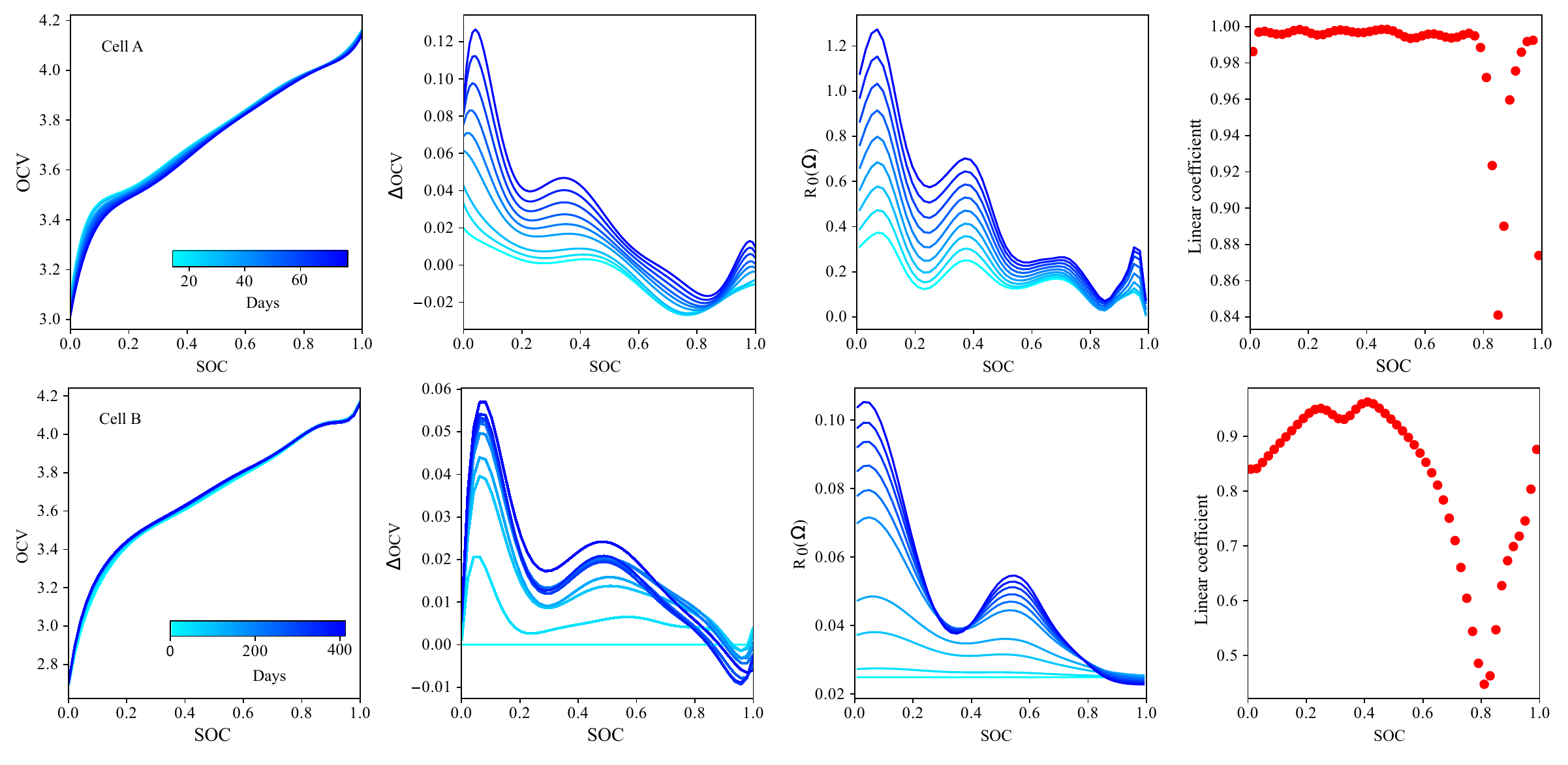}    
\caption{Comparison between ground-truth OCV-SOC and estimated $R_0$ function for cell A and B} 
\label{fig:impedance_OCV_analysis}
\end{center}
\end{figure*}

Thanks to the regular RPT data, we are able to validate this hypothesis using ground-truth OCV-SOC changes. Fig.\ \ref{fig:impedance_OCV_analysis} compares the learnt $R_0(z, \zeta)$ functions and ground-truth OCV-SOC curves. The applied current was \SI{0.125}{A} and \SI{0.34}{A} for cells A and B, respectively. Here, $\Delta \mathrm{OCV}$ represents the difference between `up-to-date' OCVs (i.e., incorporating changes as the cell ages) and BoL OCVs across all SOC levels. A strong consistency can be observed in the changing trends of $R_0$ and $\Delta \mathrm{OCV}$. This can be quantified using a Pearson linear correlation coefficient---specifically, for each discrete SOC value, the linear coefficient between $R_0$  and $\Delta \mathrm{OCV}$ as a function of age was calculated. For cell A, a linear coefficient of more than 0.8 is observed over all SOC levels, decreasing slightly around 0.8 SOC. Similarly, for cell B, a coefficient of more than 0.7 is seen for most SOC levels, decreasing around 0.8 SOC. 

One remaining issue concerns the (un)identifiability of $\Delta V_0(z, \zeta)$ and $\hat{R}_0(z, \zeta)$. Given the additive nature of these, it is impossible to distinguish them individually from terminal voltage measurements. However, the differences of their influences on terminal voltage can be compared. If one assumes, for example, a 50\% increase in resistance compared to BoL DC resistance, this would result in overpotential changes of \SI{8}{mV} and \SI{4}{mV} for cells A and B at 0.5C and 0.1C, respectively. However, the average changes in the OCV-SOC function over life are \SI{30}{mV} and \SI{20}{mV}, respectively---on the order of ten times higher.

\subsection{Differential voltage analysis}
\begin{figure*}
\begin{center}
\includegraphics[width=12cm]{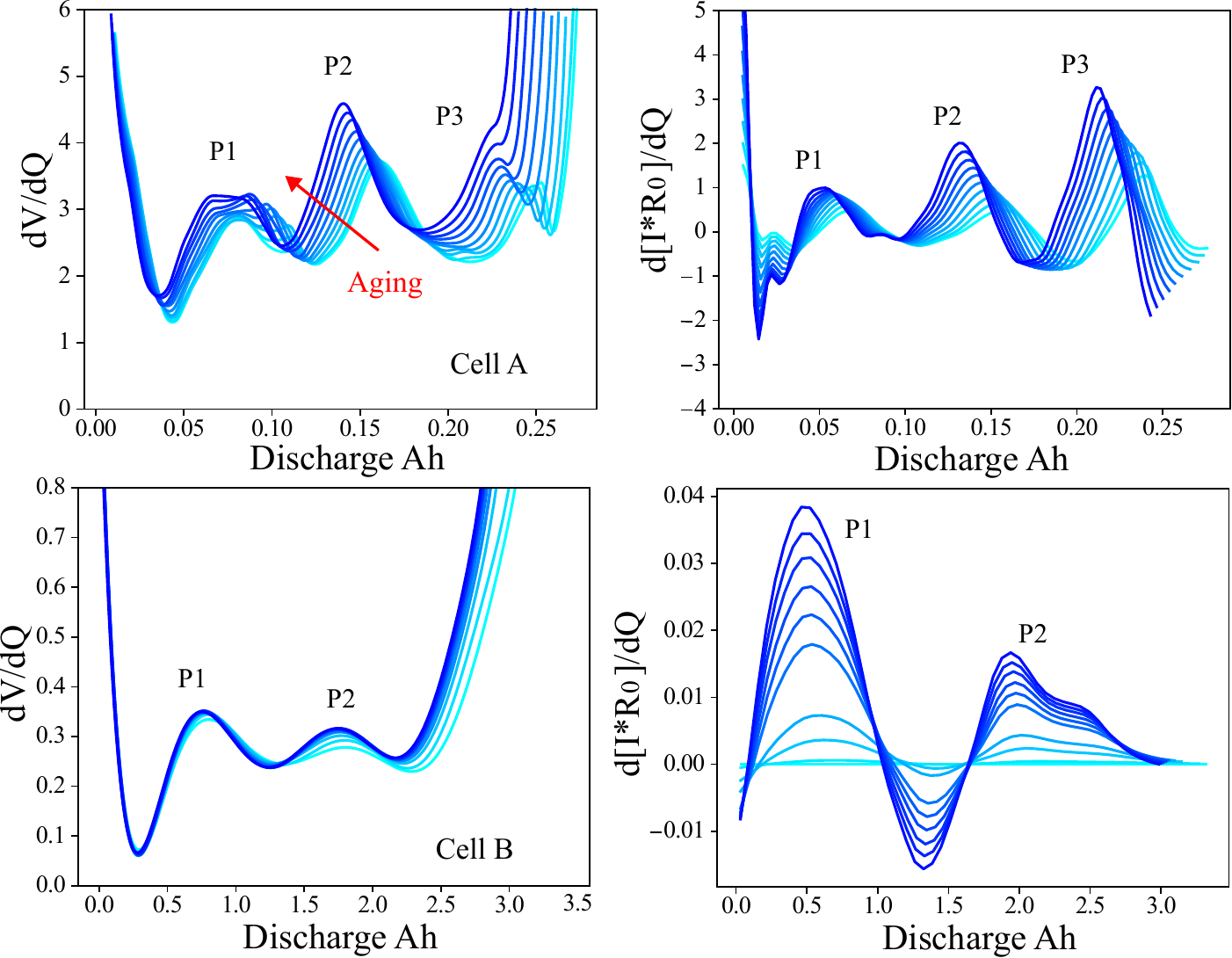}  
\caption{Comparison between RPT $dV/dQ$ and estimated $d[I  R_0]/dQ$} 
\label{fig:dqdv_combine}
\end{center}
\end{figure*}


Capacity decrease and impedance increase reflect battery SOH changes at the full cell level, but are insufficient to elucidate different degradation modes. One commonly used quantitative method for the latter is differential voltage analysis ($dV/dQ$ vs.\ discharge Ah). The $dV/dQ$ curve is obtained by differentiating discharge voltage with respect to cumulative discharge Ampere-hours. The peaks on the $dV/dQ$ curve can be assigned to electrode material phase transitions. The peak shifts and the changing between-peak distances are used to indicate different degradation modes, i.e.\ lost lithium inventory $\text{LLI}$, and loss of active material in the negative and positive electrodes, $\text{LAM}_{n}$ and $\text{LAM}_{p}$ \citep{dubarry2022best}. One major limitation of the $dV/dQ$ method is the requirement for a low C-rate ($<$C/20) deep (dis)charge, which is generally unavailable except in a carefully controlled lab environment. In our case, because of experiment limits, the RPT C-rates for cells A and B were higher, at C/5 and C/10, respectively. 

Given the observed close similarity between $\Delta \text{OCV}(z,\zeta)$ and $R_0(z, \zeta)$ as discussed in the previous section, a natural extension is to compare the $dV/dQ$ information from the RPTs with the learnt $d[I R_0]/dQ$ vs.\ discharge curve. As shown in Fig.\ \ref{fig:dqdv_combine}, if one multiplies the SOC with the learnt capacity $Q(\zeta)$, one can derive a corresponding discharge Ampere-hour estimate and thus observe similar peak shifts to differential voltage functions. 

\begin{figure*}
\begin{center}
\includegraphics[width=12cm]{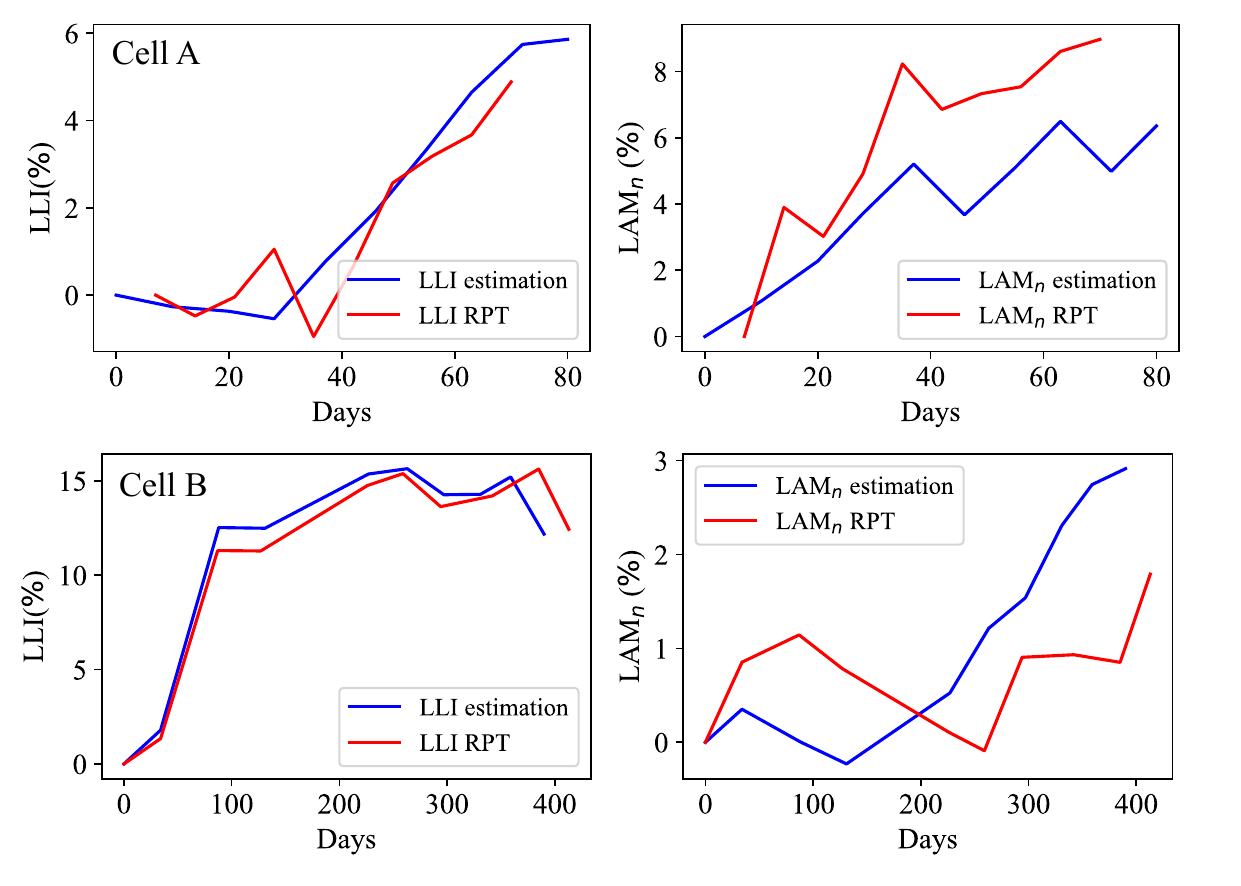}    
\caption{Comparison of estimated LLI and $\text{LAM}_n$ for $dV/dQ$ from RPTs and $d[I R_0]/dQ$ estimations.} 
\label{fig:LLI_LAM_compare}
\end{center}
\end{figure*}

Due to the lack of half cell test data, we are not able to attribute each $dV/dQ$ peak to a corresponding electrode, However, we can speculate (on the basis of literature \citep{keil2016calendar,zhu2020investigation} and our understanding for these cell chemistries) about what the peak shifts mean, and thus estimate $\text{LLI}$, $\text{LAM}_\text{n}$ for cells A and B respectively (Fig. \ref{fig:LLI_LAM_compare}). (Positive electrode loss of active material is less clear so we omit estimation of $\text{LAM}_\text{p}$.) For cell A with a CC discharging profile and NMC cathode, a 6-8 $\%$ increase is observed for both LLI and $\text{LAM}_\text{n}$. This is consistent with the relatively uniform capacity drop across life in Fig.\ref{fig:CellA_result}. For cell B, with a dynamic driving discharging profile and NCA cathode, there is a sharp increase of LLI within the first 100 days, which is likely related to diffusion-limited solid electrolyte interphase (SEI) laryer formation \citep{an2016state, spotte2022toward}. This is also consistent with capacity drop in Fig.\ref{fig:CellB_result} which then levels off.

\section{Conclusions}

We have proposed an aging-aware battery ECM incorporating GPs to describe the dependency of circuit-model parameters on SOC and lifetime. A state-space formulation was used to transform the GPs into time-stepping models enabling solution in linear time complexity with a Kalman filter and RTS smoother. This also enables co-estimation of both parameters and states in an unified framework. Compared to parameter random walks or deterministic models, using GPs to model parameter evolution has several advantages:
\begin{itemize}
  \item The state-space formulation of a GP maintains gradient information from the states whilst random walk or empirical models do not.
  \item Incorporating the SOC and aging-time dependency of parameters via GP kernels makes ECM parameter estimates more stable over battery life, i.e.\ gives smoother interpolation and more reasonable extrapolation compared to random walk or deterministic models.
  \item A GP prior helps to mitigate numerical ill-conditioning as it has a regularisation effect when parameters are not easily identifiable (e.g.\ due to measurement noise or extrapolation to unseen conditions).
\end{itemize}

We found that estimates of capacity $Q$ exhibited good consistency compared with independent ground-truth measurements from RPTs, whereas estimates of the resistance function $R_0$ showed a close relationship with shape changes in the OCV-SOC curve as a battery ages. A major part of the learnt $R_0(z, \zeta)$ function comes from OCV changes, $\Delta V_0(z, \zeta)$. High linear coefficients ($>0.8$) were observed between $R_0(z, \zeta)$ and $\Delta V_0(z, \zeta)$ over a wide range of SOCs. By using the estimates of both $Q(\zeta)$ and $R_0(z, \zeta)$, we further calculated $d[I R_0]/dQ$ as an approximation to a differential voltage $dV/dQ$ function. The estimated peak shifts and degradation modes ($\text{LLI}$, $\text{LAM}_\text{n}$) in $d[I  R_0]/dQ$ were found to show similar behaviours compared with ground-truth from $dV/dQ$.


This work highlights both a challenge, to ensure that shape changes to the OCV curve do not distort battery SOH estimates, and an opportunity---to estimate degradation modes (lost lithium inventory, lost electrode material) directly from operational data. Further tests are required on other cell form factors, duty cycles (e.g.\ shallow cycling), chemistries, temperature variations, and cell sizes to explore the generalisation of these results.

\section{Acknowledgements}
This work was supported by the Department of Engineering Science, University of Oxford, The China Scholarship Council, and Rimac Automobili.

\section{Appendix}
\appendix
\section{Gaussian process with input uncertainty}
\label{appenA}
In the joint system dynamics (eqn.\ \ref{eq:joint_sys}), $R_0$ needs to be evaluated at every time step. The SOC at step $i$ is itself a Gaussian distribution expressed as
\begin{equation}
z_i \sim \mathcal{N}(z_i^{-}, \mathbf{P}_{z_i}^{-}).
\end{equation}
This means that $R_0$ is a GP with input uncertainty from $z_i$. To incorporate this, the predictive distribution in eqn.\ \ref{eq:gp_solution} needs to be marginalised (averaged) over all $z_i$,
\begin{equation}
p\left(R_0\right)=\int p\left(R_0 \mid z_i\right) p\left(z_i\right) \mathrm{d} z_i .
\end{equation}
This integral is usually intractable but can be approximated using Taylor expansions \citep{girard2002gaussian}. A first-order expansion can be written as 
\begin{equation}
\begin{aligned}
& m_{R_0}=K_{Z, z_i}^{\top}\left[K_Z + \sigma_\text{GP}^2 I\right]^{-1} z_i \\
&\begin{split}
\mathbf{\Sigma}_{R_0}&=K_{z_i,z_i}-K_{Z,z_i}^{\top}\left[K_Z+\sigma_\text{GP}^2\right]^{-1}K_{Z,z_i}  \\
&+\left.\left.\frac{\partial R_0}{\partial z}\right|_{z_i} ^{\top} \mathbf{P}_{z_i}^{-} \frac{\partial R_0}{\partial z}\right|_{z_i}.
\end{split}
\end{aligned}
\end{equation}
where $K_Z$ and $K_{Z, z_i}$ are covariance matrix and vector following the same definition in Eqn.\ \ref{eq:gp_solution}.

\section{Predictive variances for terminal voltage}
\label{appenB}
In the extended Kalman filter used for the joint GP-ECM system, an extra variance term $I_i^{2} \mathbf{\Sigma}_{R_0, i}$ is included in the calculation of the innovation sequence $\mathbf{S}_i$. This corresponds to the contribution of the $R_0$ uncertainty when calculating the predictive distribution of terminal voltage $V_i$:
\begin{equation}
\begin{aligned}
V_i & =V_0\left(z_i\right)+I_i R_{0, i}+\nu_i, \quad \nu_i \sim N\left(0, \sigma_v^2\right) \\
\mathrm{E}\left[V_i\right] & =V_0\left(z_i\right)+I_i m_{R_{0 i}} \\
\operatorname{cov}\left(V_i\right) & =\mathrm{E}\left[\left(V_i-\mathrm{E}\left[V_i\right]\right)\left(V_i-\mathrm{E}\left[V_i\right]\right)^{\top}\right] \\
\operatorname{cov}\left(V_i\right) & =I_i^2 \Sigma_{R_{0 i}}+\sigma_\nu^2 .
\end{aligned}
\end{equation}

\section{Impedance dependency on applied current}
\label{appenC}
\begin{figure*}
\begin{center}
\includegraphics[width=16cm]{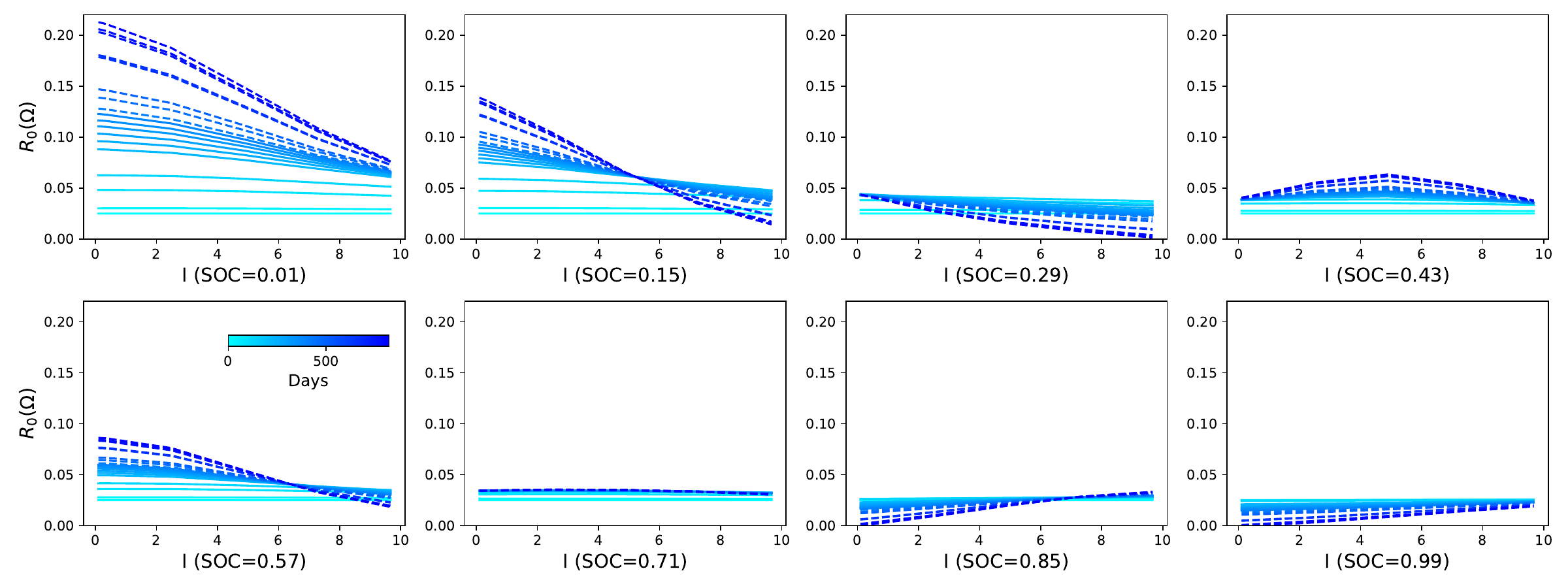} 
\caption{Dependence of resistance on applied current at several SOCs, for cell A} 
\label{fig:current_dependency}
\end{center}
\end{figure*}

Fig.\ \ref{fig:current_dependency} gives a full set of resistance estimates at several SOCs and currents, for cell A.

\section{The random-walk baseline model}
\label{appenD}
\begin{figure*}
\begin{center}
\includegraphics[width=14cm]{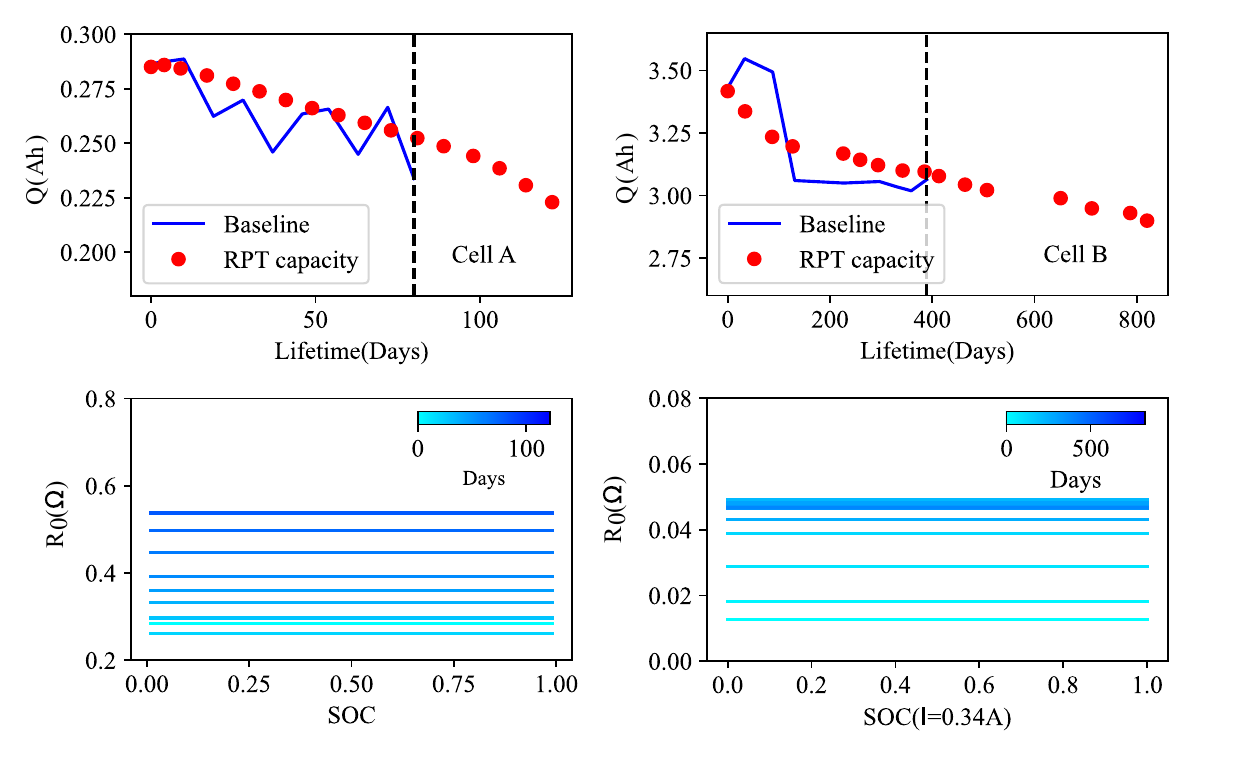} 
\caption{Results for the random-walk baseline model: cell A estimates have \SI{0.129}{Ah} RMSE and 3.27$\%$ MAPE, Cell B estimates have \SI{0.019}{Ah} RMSE and 6.74$\%$ MAPE.} 
\label{fig:baseline}
\end{center}
\end{figure*}
To benchmark the proposed method, a random-walk model was used to track aging with the commonly used dual-estimation approach \citep{plett2004extended, wei2017multi, wassiliadis2018revisiting}. The same ECM in Fig.\ \ref{fig:baseline} was used, without the dependency of $R_0$ on the operating condition.

\printcredits

\bibliographystyle{cas-model2-names}

\bibliography{cas-refs}

\end{document}